\documentstyle[12pt]{article}
\begin{document}

\def\question#1{{{\marginpar{\small \sc #1}}}}
\newcommand{\qcd'}{$\tilde{\rm QCD}$}
\newcommand{\eq}{\begin{equation}}
\newcommand{\en}{\end{equation}}
\newcommand{\bino}{\tilde{b}}
\newcommand{\tsquark}{\tilde{t}}
\newcommand{\gluino}{\tilde{g}}
\newcommand{\photino}{\tilde{\gamma}}
\newcommand{\wino}{\tilde{w}}
\newcommand{\mtilde}{\tilde{m}}
\newcommand{\higgsino}{\tilde{h}}
\newcommand{\gsi}{\,\raisebox{-0.13cm}{$\stackrel{\textstyle>}
{\textstyle\sim}$}\,}
\newcommand{\lsi}{\,\raisebox{-0.13cm}{$\stackrel{\textstyle<}
{\textstyle\sim}$}\,}

\rightline{RU-95-73}
\baselineskip=18pt
\vskip 3.5cm
\begin{center}
{\large \bf SUSY Breaking and Light Gauginos [SUSY95]}\\

\vspace*{6.0ex}

{\large Glennys R. Farrar}\footnote{Invited talk at Colloque SUSY 95,
Paris, May 1995.  Research supported in part by NSF-PHY-91-21039} \\

\vspace{1.5ex}

{\it Department of Physics and Astronomy \\ Rutgers University,
Piscataway, NJ 08855, USA}\\
\end{center}

\vspace*{4.5ex}

\centerline{\bf Abstract}

Several supersymmetry breaking mechanisms do not produce dimension-3
operators.  I show here that this scenario is consistent with present
observations and has several significant virtues: i)  When there are
no dimension-3 SUSY-breaking operators there is no SUSY-CP problem.
ii) SUSY-breaking need not occur through gauge singlets, so
that the cosmological problems often encountered in hidden sector
SUSY-breaking can be avoided. iii) Photino and gluino and $R$-hadron
masses are naturally consistent with relic photinos providing the
required dark matter density.  

Requiring spontaneous electroweak symmetry implies that scalar masses
are mostly in the $\sim 100$ GeV range.  The gluino and photino
are massless at tree level.  At 1-loop, the gluino and photino masses
at the ew scale are predicted to be $m_{\gluino}\sim 10 - 600$ MeV
and $m_{\photino} \sim 100 - 1400$ MeV.  New hadrons
with mass $\sim 1 \frac{1}{2}$ GeV are predicted and described.  The
``extra'' flavor singlet pseudoscalar in the $\iota(1440)$ region
which has been observed in several experiments is naturally
interpreted as the mainly-$\gluino \gluino$ bound state which gets its
mass via the QCD anomaly.  Its superpartner, a gluon-gluino bound
state, would be the lightest $R$-hadron.  For the most interesting
portions of parameter space the $R^0$ lifetime is $10^{-6} - 10^{-10}$
sec, so existing searches would not have been sensitive to it.  Search
strategies and other consequences of the scenario are briefly
mentioned. 

\newpage

The customary approach to studying the phenomenological implications
of supersymmetry has been to assume that the ``low energy'' effective
Lagrangian below the SUSY-breaking scale, $M_{SUSY}$, contains all
possible renormalizable operators, including in principle all possible
soft supersymmetry breaking terms, consistent with the gauge
symmetries and possibly some global and discrete symmetries.  Some
models of SUSY-breaking naturally lead to relations among the SUSY-breaking
parameters at the scale $M_{SUSY}$ so that the minimal supersymmetric
standard model (MSSM) requires specification of 6-8 parameters beyond
the gauge and Yukawa couplings already determined in the MSM: $tan
\beta \equiv \frac{v_U}{v_D}$, the ratio of the two Higgs vevs, $\mu$,
the coefficient of the SUSY-invariant coupling between higgsinos,
$M_0$, a universal SUSY-breaking scalar mass, $m_{12}^2$, the
SUSY-breaking mixing in the mass-squared matrix of the Higgs scalars
(aka $\mu B$ or $\mu M_0 B$ in alternate notations), $M_{1,2,3}$, the
SUSY-breaking gaugino masses (proportional to one another if the MSSM
is embedded in a GUT), and $A$, the coefficient of SUSY-breaking terms
obtained by replacing the fermions in the MSM Yukawa terms by their
superpartners.  To obtain predictions for the actual superparticle
spectrum in terms of these basic parameters, the renormalization group
equations for masses, mixings and couplings are evolved from the scale
$M_{SUSY}$ to the scale $M_{Z^0}$ where, on account of different RG
running and flavor dependent couplings, the various scalars and
fermions have quite different masses.  A particularly attractive
aspect of this approach is that for the heavy top quark which is found
in nature, the mass-squared of a combination of Higgs fields becomes
negative at low energy and the electroweak symmetry is spontaneously
broken\cite{ir,a-gpw}, with $m_Z$ a function of $A,~ M_0$ and other
parameters of the theory. In this conventional treatment of the MSSM,
squark and gluino masses are constrained by experiment to be greater
than $\sim 100$ GeV\cite{cdf:gluinolim2}. 

I will argue here that a more restrictive form of SUSY breaking, one
without dimension-3 operators, is actually superior in several
respects.  We shall see that the remaining parameters of the theory
are well-constrained when electroweak symmetry breaking is demanded,
and that the resultant model is both extremely predictive and
consistent with laboratory and cosmological observations.  If this is
the correct structure of the low energy world, there will be many
consequences which can be discovered and investigated before the
construction of the LHC.  Some of these are discussed below.

There are at least two good reasons for dispensing with dimension-3
SUSY breaking operators in the low energy theory.  First, in order to
produce dimension-3 operators in the appealing
hidden-sector-dynamical-SUSY-breaking scenarios, the auxilliary
component of one or more hidden sector {\it gauge singlet} field
must develop a large vev.  However as was emphasized in
ref. \cite{bkn}, this implies particles with masses in the
100 GeV - 1 TeV region which are in conflict with cosmology, in
particular causing late-time entropy production which is incompatible
with primordial nucleosynthesis.

Besides avoiding the problems associated with singlet fields, not
having dimension-3 SUSY breaking is attractive because it solves the
SUSY CP problem.  This can be seen in the MSSM as follows.  In the
absence of dimension-3 SUSY-breaking operators, the only phases beyond
those of the MSM (the phases in the quark mass matrix and the $\theta$
parameter) which can be present in the tree level theory appear in the
parameters $\mu$ and $m_{12}^2$. However a combination of an
$R$-transformation and U(1) transformations on the Higgs 
superfields allows these phases to be removed.  Any phase which is
introduced thereby into the Yukawa terms in the superpotential can be
removed by chiral transformations on the quark superfields, merely
changing the phases which contribute to the strong CP problem (which
must be solved by some other mechanism).  Since the gauge-kinetic
terms are not affected by U(1) and $R$ transformations, the preceding
manipulations do not introduce phases in interactions involving
gauginos\footnote{See ref. \cite{f:101} for the more general case when
the squark mass-squared is off-diagonal and a discussion of other
related mechanisms.}.     

When there are no dimension-3 SUSY-breaking operators, $A$ and
$M_{1,2,3}$ are zero.  The gluino and lightest neutralino are massless
in tree approximation but they get masses at one loop from virtual
quark-squark pairs.  The neutralinos also get a contribution from
``electroweak'' loops involving wino/higgsino-Higgs/vector boson
pairs\cite{bgm,pierce_papa,f:96}.  The sizes of these corrections were
estimated, for various values of $M_0, ~\mu$, and $ tan \beta$, in
ref. \cite{f:96}.  There, it was determined that in order to insure
that the chargino mass is greater than its LEP lower bound of about 45
GeV, $\mu$ must either be less than 100 GeV (and $tan \beta \lsi 2$)
or greater than several TeV.  Here I will also demand that the
electroweak symmetry breaking produces the observed $m_Z$ for
$m_{t}\sim 175$ GeV.  This is not possible in the large $\mu$ region,
so I will consider only $\mu \lsi 100$ GeV.  In addition, from Fig. 6
of ref. \cite{a-gpw} one sees that $M_0$, the SUSY-breaking scalar
mass, must be $\sim 100 - 300 $ GeV, with 150 GeV being the favored
value\footnote{As shown by Lopez\cite{lopez} et al and
Diaz\cite{diaz}, the requirement of GUT unification and a {\it single,
universal} SUSY-breaking scalar mass at the SUSY-breaking scale is at
best marginally compatible with the chargino, neutralino and Higgs
mass bounds.  The deviation from universality need not be great to
avoid this problem, however.}.  From Figs. 4 and 5 of ref. \cite{f:96}
this gives $m_{\gluino} \sim 100-600$ and $m_{\photino} \sim 100-900$
MeV at the ew scale. Since the electroweak loop was treated in ref.
\cite{f:96} with an approximation which is valid when $M_0$ or $\mu$
is $>> m_Z$, those results for the photino mass are only indicative of
the range to be expected.  Until a more precise calculation is
available, we attach a $\sim$factor-of-two uncertainty to the
electroweak loop, and consider the enlarged photino mass range $100 -
1400$ MeV. 

The most essential aspects of the phenomenology of this theory are: 
\begin{enumerate}

\item  Predicted mass and lifetime of the lightest $R$-meson, the $g
\gluino$ bound state denoted $R^0$.

\item  Predicted mass of the flavor singlet pseudoscalar which gets
its mass via the anomaly (the ``extra'' pseudoscalar corresponding to
the $ \gluino \gluino$ ground state degree of freedom).   

\item  Identity of the flavor singlet pseudogoldstone boson resulting
from the spontaneous breaking of the extra chiral symmetry associated
with the light gluino. 

\item  Properties of the flavor-singlet $R$-baryon composed of
$uds\gluino$, called $S^0$. 

\item  Production rates and detection strategies for the new $R$-hadrons.

\end{enumerate}

In order to estimate the $R^0$ lifetime, we need its mass.
Fortunately, it can be quite well determined from existing lattice QCD
calculations, as follows\cite{f:95}. If the gluino were massless and
there were no quarks in the theory (let us call this theory sYM),
SUSY would be unbroken and the $R^0$ would be in a degenerate
supermultiplet with the $0^{++}$ glueball, $G$, and a $0^{-+}$ state
denoted $ \tilde{\eta}$, which can be thought of as a $\gluino 
\gluino$ bound state. To the extent that quenched approximation is
accurate for sYM\footnote{The 1-loop beta function is the same for sYM
as for ordinary QCD with 3 light quarks, so the accuracy estimate for 
quenched approximation in ordinary QCD, $10-15\%$, may be applicable
here.}, the mass of the physical $R^0$ in the continuum limit of this
theory would be the same as the mass of the $0^{++}$ glueball.  The
latest quenched lattice QCD value of $m(G)$ from the GF11 group is
$1740 \pm 71$ MeV\cite{weingarten:glu1740}\footnote{Note the increase
from the $1440 \pm 110$ value given in ref.
\cite{weingarten:glueballs} and used in my SUSY95 talk and earlier 
work\cite{f:95,f:99}.}.  The UKQCD collaboration
reports\cite{ukqcd:glueballs} $1550 \pm 50$ MeV for the $0^{++}$ mass,
but this error is only statistical.  Adding in quadrature a $70$ MeV
lattice error and a 15\% quenching uncertainty\footnote{The
uncertainty associated with quenched approximation with both light
quarks and gluinos was taken in \cite{f:95} to be 25\%.  However since
the estimate of the quenching error for ordinary QCD is obtained by
comparing lattice results with the hadron spectrum, it already
includes the effects of gluinos, if they are present in nature.}
leads to a total uncertainty of $\sim 270$ MeV, so I will use the
range 1.3 - 2 GeV for massless gluinos.  Experimentally, the
$f_0(1520)$ and $f_0(1720)$ seem to be the leading candidates for the
ground state glueball, but the situation is still
unclear\cite{amsler_close}.  Mixing with other flavor singlet
pseudoscalars can shift the $\tilde{\eta}$ somewhat.  For gluino
masses small compared to the ``confinement mass'' of $\sim 1
\frac{1}{2}$ GeV, one would expect the $R^0$ and $\tilde{\eta}$ masses
to be insensitive to the gluino mass.  Thus in the absence of a
dedicated lattice gauge theory calculation of the masses of these
particles, we can adopt the estimate, $1.3 - 2$ GeV for all these
states, $G,~\tilde{\eta}$, and $R^0$. 

This discussion shows that in sYM, which is identical to ordinary QCD
in quenched approximation, the $\tilde{\eta}$ with mass $\sim 1
\frac{1}{2}$ GeV is the pseudoscalar which gets its mass from the
anomaly.  Thus in QCD with light gluinos the particle which gets its
mass from the anomaly is predicted to be too heavy to be the $\eta'$.
Instead, the $\eta'$ should be identified with the pseudogoldstone
boson associated with the spontaneous breaking by quark and gluino
condensates of the non-anomalous linear combination of the light
quarks' chiral $U(1)$ and the gluinos' chiral $U(1)_R$
symmetries\cite{f:95}.  Using standard PCAC and current algebra
techniques, I obtained in ref. \cite{f:95} the relationship between
masses and condensates necessary to produce the correct $\eta'$ mass
(ignoring mixing): $m_{\gluino}$ (at the hadronic scale) times $ <
\bar{\lambda}\lambda>~ \sim 10~ m_s <\bar{s}s>$.  The required gluino
condensate is reasonable, for $m_{\gluino} \gsi 100$ MeV.\footnote{The
top-stop loop produces $m_{\gluino} \gsi 100$ MeV at the ew scale as
long as $\mu cot \beta$ is not too small and $M_0 \lsi 300$
GeV\cite{f:96}.  Since the mass at the hadronic scale is larger than
that at the ew scale, it is not difficult to accomodate this
inequality.} In a more refined discussion, the physical $\eta'$ would
be treated as a superposition of the pseudo-goldstone boson and the
orthogonal state which gets its mass from the anomaly.  Surprising as
it may seem, given present experimental and theoretical uncertainties,
the $\eta'$ could be this pseudogoldstone boson, containing $ \sim 30 \%$
$\gluino \gluino$ component.  This is because accurately-tested,
model-independent predictions concerning the $\eta'$ are for ratios in
which the gluino component plays no role.

Note parenthetically that this scenario {\it predicts} the existance
of a flavor singlet pseudoscalar meson {\it in addition} to the
$\eta'$, which is not a part of the conventional QCD spectrum of quark
mesons and glueballs, whose mass should be $\sim 1.3 - 2$ GeV.  Such a 
state has been found at $\sim 1410-1440$ MeV in radiative $J/\Psi$
decay\cite{dm2,mark3} and, since SUSY95, in $p \bar{p}$
annihilation\cite{amsler:E}.  This state is extremely difficult to
accomodate in conventional QCD since:
\begin{enumerate}
\item The first excited pseudoscalar nonet is filled by
$\pi{1300},~K{1460},~\eta{1295}$ and $\eta{1440}$.  The non-singlet
members of the second excited pseudoscalar nonet are also clear:  the
$\pi{1770}$ and $K{1830}$.  A flavor singlet member of this nonet
cannot plausibly be as light as $\sim 1420$ MeV, and moreover suitable
candidates for the two isosinglet members of the nonet are available.
\item  Experimental evidence (and also lattice QCD) points to the
scalar glueball being in the 1500-1700 MeV region. Since the
pseudoscalar glueball is an $L=1$ state, a naive analysis would place
it higher in mass than the scalar glueball which has $L=0$.  This
simple intuition is confirmed by lattice QCD
calculations\cite{ukqcd:glueballs} which indicate the $0^{-+}$
glueball should be $\sim \frac{1}{2}$ GeV heavier than the $0^{++}$
glueball.  Thus 1420 MeV appears to be an unreasonable mass for a
pseudoscalar glueball.  
\item  The only remaining possibility advanced within conventional QCD
for a state which is neither glueball nor a member of a nonet, is that
it is a weakly bound ``molecule'' which therefore appears near
threshold for production of the components of the molecule.  Since the
$K-K^*$ threshold is at 1390 MeV, this might be an explanation.  But
since the extra state under discussion is a pseudoscalar, the
molecule would have to have $L=1$.  Due to the angular momentum
barrier, binding in this channel seems extremely unlikely.  Moreover
if the $L=1$ channel binds, one would expect much stronger binding in
the $L=0$ channel and such states have not been observed.
\item Other aspects of the state such as its width and production rate
in radiative $J/\Psi$ decay are consistent with its being primarily
composed of gluons or gluinos and not compatible with its containing
quarks\cite{f:93}.    
\end{enumerate}
 
For many years the famous UA1 figure\cite{ua1} and its decendants,
showing the allowed regions of the gluino-squark mass plane, has
widely been accepted as excluding all but certain small ``windows''
for low gluino mass.  However there are two significant problems with
that analysis which makes it not valid for the very light gluino
region.  First off all, no distinction was made between the mass of
the gluino and the mass of the lightest hadron containing it.
Although this distinction is unimportant for heavy gluinos, it is
essential when the gluino is light.  Since the $R^0$ mass is $\sim 1
\frac{1}{2}$ GeV in the massless gluino limit, this distinction is
even more important for hadrons containing gluinos than it is for
hadrons composed of quarks.  Perturbative QCD predictions miss the
kinematic suppression associated with non-perturbative mass-generation
of the actual $R$-hadrons being produced.  This results in a some
cases in a serious overestimate of the expected production rate, and
thus an exagerated view of the experimental sensitivity.  Papers
relying on perturbative QCD typically report an excluded range of
gluino mass, but for a very light gluino it would actually be closer
to the truth to replace their ``gluino mass'' with half the mass of
the $R^0$. 

The second and more serious problem with the UA1 compilation is that it
{\it assumed} that the lightest $R$-hdaron's lifetime is short enough
for missing energy and beam dump experiments to be sensitive to the
neutralino produced by the $R$-hadron decay.  However $R$-hadrons
produced in the target or beam dump degrade in energy very rapidly due
to their strong interactions.  Thus only when the photino is emitted
before the $R$-hadron interacts, will it have enough energy to pass
the missing energy cut or be recognized in the downstream detector.
As discussed in connection with a particular experiment in ref.
\cite{f:55}, and more generally in ref. \cite{f:95}, if the $R^0$
lifetime is longer than $\sim 5~10^{-11}$ sec this criterion is not
met and beam dump and missing energy experiments become increasingly
``blind'' to light gluinos as their lifetime increases.  
  
Therefore we must estimate the $R^0$ lifetime.  Making a
first-principles estimate of the lifetime of a light hadron is always
problematic.  Although the relevant short distance operators can be
accurately fixed in terms of the parameters of the Lagrangian,
hadronic matrix elements are difficult to determine.  It is
particularly tricky for the $R^0$ in this scenario because the photino
mass is larger than the current gluino mass and, since $m_{\photino}
\sim \frac{1}{2} m_{R^0}$, the decay is highly suppressed even using a
constituent mass for the gluino.  The decay rate of a free gluino into
a photino and massless $u \bar{u}$ and $d \bar{d}$ pairs is well
known.  The problem is to take into account how interactions with the
gluon and ``sea'' inside the $R^0$ ``loans'' mass to the gluino.  If
this effect is ignored one would find the $R^0$ to be absolutely
stable except in the upper portion of its estimated mass range.

A method of estimating the {\it maximal} effect of such a ``loan'',
and thus a lower limit on the $R^0$ lifetime, was obtained in ref.
\cite{f:99} by elaborating a suggestion of refs. \cite{accmm,franco}.
The basic idea is to think of the hadron as a bare massless parton (in
this case a gluon) carrying momentum fraction $x$ and a remainder
(here, the gluino) having an effective mass $M\sqrt{1-x}$, where $M$
is the mass of the decaying hadron, here the $R^0$.  Then the
structure function, giving the probability distribution of partons of
fraction $x$, also gives the distribution of effective masses for the
remainder (here, the gluino).  The distribution function of
the gluon in the $R^0$ is unknown, but can be bracketed with extreme
cases: the non-relativistic $F_{nr}(x) = \delta(x-\frac{1}{2})$ and
the ultrarelativistic $F_{ur}(x)= 6x(1-x)$.  The normalizations are
chosen so that half the $R^0$'s momentum is carried by gluons.  In the
cosmologically interesing region (see below) $1.6 \le r \le 2$, the
lifetime given by this model is between $10^{-6}$ 
and $10^{-9}$ sec for $m(R^0)=1.5$ GeV and a 150 GeV squark.

The decay rates produced in this model can be considered upper limits
on the actual decay rate, because the model maximizes
the ``loan'' in dynamical mass which can be made by the gluons to the
gluino.  For kaon semileptonic decay (where the $K_{\mu 3}$ mode would
be strongly suppressed or excluded in the absence of similar effects,
since the strange quark current mass is of the same order as the muon
mass) this model gives the correct ratio between $K_{\mu 3}$ and $K_{e
3}$ rates, and rates 2-4 times larger than observed: overestimating
the rate as we anticipated, but not by a terribly large factor. 
Although a large range of uncertainty should be attached to the $R^0$
lifetime estimated this way, these results are still useful because
they give lower bounds on the lifetime.  Even with the
ultrarelativistic wavefunction which gives the shortest lifetime
estimate, for most of the parameter space of interest the lifetime is
long enough that energy degradation in a beam dump experiment is
significant.  See ref. \cite{f:101} for a detailed discussion and
revised constraints. 

I explained above why the usual beam dump and missing energy limits
are not applicable to the very-light, long-lived gluinos predicted in
this scenario.  In refs. \cite{f:95,f:101} I also investigated other
experiments which are potentially relevant and found that the
interesting regions of mass ($m(R^0) \lsi 2 $ GeV) and lifetime
($\tau(R^0) \gsi 10^{-10} s$), are essentially unconstrained
experimentally!  However photinos of mass $\lsi 1$ GeV are an integral
prediction of this scenario and such light photinos are commonly
believed to be excluded.  As was the case for the light gluinos, 
closer examination reveals that this is not actually correct when the
details of the scenario are taken into account.  First of all, since
gaugino masses come from radiative corrections, limits relying on GUT
tree-level relations between gaugino masses do not apply.
Furthermore, the lightest and next-to-lightest neutralinos (called in
general $\chi^0_1$ and $\chi^0_2$) are not produced in $Z$ decays with
sufficient rate to be observed at LEP, because in this scenario the
$\chi^0_1$ is extremely close to being pure photino\cite{f:96}.  It
contains so little higgsino that $Z^0 \rightarrow \chi_1^0 \chi_1^0$
and $Z^0 \rightarrow\chi_1^0 \chi_2^0$ are suppressed compared to the 
conventional scenario with tree-level gaugino masses.  Except for
very small $\mu$, the mass of the $\chi^0_2$ is high enough
that its pair production is too small to be important for most of
parameter space, since $tan \beta$ is already required to be fairly
small from the chargino mass limit.

It has also been claimed that a stable photino with mass less than
$\sim 15$ GeV is excluded because it would produce too large a relic
abundance, ``overclosing'' the universe.  These calculations assumed
that self annihilation was the only important mechanism for keeping
photinos in thermal equilibrium, leading to freeze out at a
temperature $T \sim \frac{1}{14} m_{\photino}$, below which the
self-annihilation rate is less than the expansion rate of the
universe.  However it has recently been shown\cite{f:100} that when
the gluino is also light, other processes involving $\photino - R^0$
interconversion followed by $R^0$ self-annihilation become important,
and freezeout is delayed to a lower temperature.  When the ratio $r =
\frac{m(R^0)}{m_{\photino}}$ is in the range 1.6 - 2, the relic
photino abundance can account for the dark matter of the
universe\cite{f:100}.  Only when $r \gsi 2.2$ would the photino relic
density be unacceptably large.  If that were the case, this scenario
would only be viable if $R$-parity were violated, so that the photino
was not absolutely stable.  Note that the desirable range $1.6 < r <
2.2$ is consistent with the $R^0$ and photino masses expected in this
scenario (even predicted in advance of the dark matter density
calculation of ref. \cite{f:100}).

There is another interesting light $R$-hadron besides the $R^0$,
namely the flavor singlet scalar baryon $uds \gluino$ denoted $S^0$.
In view of the very strong hyperfine attraction among the
quarks\cite{f:51}, this state may be similar in mass to the
$R^0$.\footnote{It is amusing that the (spin 1/2) baryon spectrum
contains an anomalous state, the $\Lambda(1405)$, which this
discussion suggests is likely to be a $uds$-gluon flavor singlet
cryptoexotic baryon.  The validity of this suggestion for identity of
the $\Lambda(1405)$ is of course independent of the existance of light
gluinos.}  If its mass is $\sim 1 \frac{1}{2}$ GeV, it will be
extremely long lived, or even stable if $m(S^0) - m(p) - m(e^-) <
m_{\photino}$.  Even if decay to a photino and nucleon is
kinematically allowed, the decay rate will be very small since it
requires a flavor-changing-neutral transition as well as an
electromagnetic interaction.  If the $S^0$ does not bind to nuclei,
its being absolutely stable is not experimentally
excluded\cite{f:51,f:95}.  In fact the $S^0$ seems unlikely to bind to
nuclei since the two-pion-exchange force, which is attractive
between nucleons but insufficient to explain their binding, is
repulsive in this case\cite{f:95} because the mass of the intermediate
$R_{\Lambda}$ or $R_{\Sigma}$ is much larger than that of the
$S^0$.\footnote{If its decay to a nucleon is kinematically forbidden,
the $S^0$ would be absolutely stable unless both $R$-parity and
baryon-number conservation are violated.  In several models of $R$-parity
violation, e.g. \cite{masiero_valle}, $R$-parity is violated in
association with lepton number violation, but baryon number is
conserved.  In this case the photino would be unstable (e.g.,
$\photino \rightarrow \nu \gamma$) but the $S^0$ stable.  Stable relic
$S^0$'s could make up part of the missing mass in our galaxy, but
since they have strong interactions they might clump too much to
account for the bulk of the missing dark matter of the universe.}  For
further discussion of the $S^0$ and other $R$-hadrons see refs.
\cite{f:51} and \cite{f:95}.  
 
In ref. \cite{f:95} I discussed strategies for detecting or excluding
the existance of an $R^0$ with a lifetime too long to be detected via 
decays in an accelerator experiment.  If instead the $R^0$ lifetime is
in the range $\sim {\rm few} 10^{-6} - {\rm few} \times 10^{-10}$s, it
will be possible with planned rare kaon decay and $\epsilon'/\epsilon$
experiments to find evidence for the $R^0$\cite{f:102}.  If $R^0$'s
exist in the $\lsi 2 $ GeV mass range, the beam for a kaon experiment
will contain $R^0$'s, whose decays one wants to observe.  Given 
the expected $R^0$ mass and assuming the photino mass corresponds to
the cosmologically interesting region of $r$, the signal for an $R^0$
decay in the detector is distinctive: a single $\pi^0$ or $\eta$ (or
$\pi^+ \pi^-$ pair with invariant mass greater than $m_K$), with
substantial $p_{\perp}$.  See refs. \cite{f:99,f:102} for more detailed
rate estimates and further discussion. 

In summary, I have shown that when SUSY breaking only arises through
dimension-2 operators, there is no SUSY CP problem.  Gauge singlet
fields breaking SUSY are unnecessary, so that the attendant cosmological
problems are avoided.  Paramters of the theory can be constrained by
requiring correct electroweak symmetry breaking.  The lightest SUSY
particles are the photino and glueballino with masses predicted to be
in the ranges $100-1400$ MeV and $1.3-2$ GeV respectively.  For
squarks masses in the normally expected range, the lifetime
and properties of the new hadrons are not in conflict with existing
experimental limits.  Possibilities for their detection were
mentioned.  The photino is an attractive cold dark matter candidate
because, when the ratio $m(R^0)/m_{\photino}$ lies in the range $\sim
1.6 - 2$, its relic abundance leads to $\Omega h^2 \sim 0.25 - 1$
consistent with observation.  The lightest chargino must be lighter
than the $W$.   

{\bf Acknowledgements:}  I am indebted to W. Willis and M.
Schwartz for discussions of search strategies, and to collaborators in
investigating various aspects of this problem: E. W. Kolb, C. Kolda,
M. Luty, A. Masiero and S. Thomas. 




\begin{thebibliography}{10}

\bibitem{ir}
L.~Ibanez and G.~Ross.
\newblock {\em Phys. Lett.}, B110:215, 1982.

\bibitem{a-gpw}
L.~Alvarez-Gaume, J.~Polchinski, and M.~Wise.
\newblock {\em Nucl. Phys.}, B221:495, 1983.

\bibitem{cdf:gluinolim2}
F.~Abe et~al.
\newblock {\em Phys. Rev. Lett.}, 69:3439, 1992.

\bibitem{bkn}
T.~Banks, D.~Kaplan, and A.~Nelson.
\newblock {\em Phys. Rev.}, D49:779, 1994.

\bibitem{f:101}
G.~R. Farrar.
\newblock Technical Report RU-95-25 and hep-ph/9508291, Rutgers Univ., 1995.

\bibitem{bgm}
R.~Barbieri, L.~Girardello, and A.~Masiero.
\newblock {\em Phys. Lett.}, B127:429, 1983.

\bibitem{pierce_papa}
D.~Pierce and A.~Papadopoulos.
\newblock Technical Report JHU-TIPAC-940001, PURD-TH-94-04,hep-ph/9403240,
  Johns Hopkins and Purdue, 1994.

\bibitem{f:96}
G.~R. Farrar and A.~Masiero.
\newblock Technical Report RU-94-38, hep-ph/9410401, Rutgers Univ., 1994.

\bibitem{lopez}
J.~Lopez and D.~Nanopoulos nad X.~Wang.
\newblock {\em Phys. Lett.}, B313:241, 1993.

\bibitem{diaz}
M.~Diaz.
\newblock {\em Phys. Rev. Lett.}, 73:2409, 1994.

\bibitem{f:95}
G.~R. Farrar.
\newblock {\em Phys. Rev.}, D51:3904, 1994.

\bibitem{weingarten:glu1740}
H.~Chen, J.~Sexton, A.~Vaccarino, and D.~Weingarten.
\newblock {\em Nucl. Phys.}, B(Proc. Supp.)34:357, 1994.

\bibitem{weingarten:glueballs}
H.~Chen, J.~Sexton, A.~Vaccarino, and D.~Weingarten.
\newblock Technical Report IBM-HET-94-1, IBM Watson Labs, 1994.

\bibitem{f:99}
G.~R. Farrar.
\newblock Technical Report RU-95-17 and hep-ph/9504295, Rutgers Univ., April,
  1995.

\bibitem{ukqcd:glueballs}
G.~Bali et~al.
\newblock {\em Phys. Lett.}, B309:378, 1993.

\bibitem{amsler_close}
C.~Amsler and F.~Close.
\newblock Technical Report RAL-95-036; hep-ph/9505219, Rutherford Lab, 1995.
\newblock Evidence for a Scalar Glueball.

\bibitem{dm2}
J.~E. Augustin et~al.
\newblock Technical Report 90-53, LAL, 1990.

\bibitem{mark3}
Z.Bai et~al.
\newblock {\em Phys. Rev. Lett.}, 65:2507, 1990.

\bibitem{amsler:E}
The Crystal~Barrel Collaboration.
\newblock Technical Report to be published in Phys. Lett. B, CERN, 1995.

\bibitem{f:93}
M.~Cakir and G.~R. Farrar.
\newblock {\em Phys. Rev.}, D50:3268, 1994.

\bibitem{ua1}
UA1 Collaboration.
\newblock {\em Phys. Lett.}, 198B:261, 1987.

\bibitem{f:55}
G.~R. Farrar.
\newblock {\em Phys. Rev. Lett.}, 55:895, 1985.

\bibitem{accmm}
G.~Altarelli, N.~Cabibbo, G.~Corbo, L.~Maiani, and G.~Martinelli.
\newblock {\em Nucl. Phys.}, B208:365, 1982.

\bibitem{franco}
E.~Franco.
\newblock {\em Phys. Lett.}, 124B:271, 1983.

\bibitem{f:100}
G.~R. Farrar and E.~W. Kolb.
\newblock Technical Report RU-95-18 and astro-ph/9504081, Rutgers Univ., 1995.

\bibitem{f:51}
G.~R. Farrar.
\newblock {\em Phys. Rev. Lett.}, 53:1029--1033, 1984.

\bibitem{masiero_valle}
A.~Masiero and J.~Valle.
\newblock {\em Phys. Lett.}, B251:273, 1990.

\bibitem{f:102}
G.~R. Farrar.
\newblock Technical Report RU-95-26 and hep-ph/9508292, Rutgers Univ., 1995.

\end{thebibliography}

\end{document}